\documentclass[12pt,a4paper]{article}
\usepackage{graphicx}
\usepackage{times}
\title{\bf The diversity of AGN variability: Some highlights and challenges}

\author{Gopal-Krishna$^1$\thanks{gopaltani@gmail.com} , Silke Britzen$^2$, Paul Wiita$^3$\\
\vspace{0.5cm}\\
\normalsize $^1$ ARIES, Manora Peak, Nainital- 263002, India\\ 
\normalsize $^2$ MPIfR, Auf dem Huegel 69, D-53121, Bonn, Germany\\
\normalsize $^3$ Dept. of Physics, TCNJ, Ewing, NJ 08628-0718, USA}
\date{\mbox{}}
\begin{document}
\maketitle
\setcounter{page}{1001}
\pagestyle{plain}
    \makeatletter
    \renewcommand*{\pagenumbering}[1]{%
       \gdef\thepage{\csname @#1\endcsname\c@page}%
    }
    \makeatother
\pagenumbering{arabic}

%
%
\def\bull{\vrule height .9ex width .8ex depth -.1ex}
\makeatletter
\def\ps@plain{\let\@mkboth\gobbletwo
\def\@oddhead{}\def\@oddfoot{\hfil\scriptsize\bull\quad
"2nd Belgo-Indian Network for Astronomy \& astrophysics (BINA) workshop'', held in Brussels (Belgium), 9-12 October 2018 \quad\bull}%
\def\@evenhead{}\let\@evenfoot\@oddfoot}
\makeatother
%
%
\def\beginrefer{\section*{References}%
\begin{quotation}\mbox{}\par}
\def\refer#1\par{{\setlength{\parindent}{-\leftmargin}\indent#1\par}}
\def\endrefer{\end{quotation}}
%
%

{\noindent\small{\bf Abstract:}

This article focuses on certain variability and emission characteristics of Active Galactic Nuclei (AGN), 
especially their radio-loud subset consisting of quasars, BL Lacs and $\gamma$-ray detected 
narrow-line Seyfert 1 galaxies, all of which exhibit relativistically beamed jets of nonthermal radiation. 
Several striking trends and correlations, including some that have received scant attention, drawn from 
the available comparatively recent literature are highlighted. These can provide very useful inputs to models 
of AGN and be probed at a deeper level using the optical telescopes recently set up at Devasthal (Nainital).
}               
\vspace{0.5cm}

\section{Introduction}

Temporal variability of flux density across the observable electromagnetic spectrum is an outstanding 
feature of Active Galactic Nuclei (AGN), most of which are radio-quiet. Pronounced variability 
is exhibited by the tiny subset of AGN called `blazars'; they are essentially always radio loud and
their emission is predominantly Doppler boosted non-thermal radiation arising from a relativistically 
advancing jet pointed close to our direction (see, e.g., reviews by Wagner \& Witzel 1995; Marscher 2016, 
Aharonian 2017). Blazars are sub-divided into two major classes: `Flat-Spectrum Radio Quasars' (FSRQs) 
and `BL Lacs' (BLs), depending on the prominence of (broad) emission lines in the optical/UV spectrum
(e.g., Begelman et al.\ 1984; Urry \& Padovani 1995; Antonucci 2012). The jet boosting, a hallmark of 
blazars, is also shared by the small, radio-loud subset of `Narrow-Line Seyfert 1' (NLSy1) galaxies 
(e.g., Komossa et al.\ 2006; Zhou et al.\ 2007; Angelakis et al.\ 2015). The observed short time scales of 
variability (even minute like) can yield fairly tight constraints on the size, locations and kinematics 
of the emitting regions (which are far too small for direct imaging) and even about the magnetic field in 
them (e.g., Narayan \& Piran 2012; Hagen-Thorn et al. 2008). Extensive literature on rapid flux variability 
of blazars in the optical/near-infrared band, often called `Intra-Night Optical Variability' (INOV, 
Gopal-Krishna et al.\ 2003), has recently been summarized in Gopal-Krishna \& Wiita (2018) in relation 
to the ARIES program. 
Here we discuss only a selection of relatively recent observational findings that have either clarified 
some much debated issues about blazars/AGN, or posed some intriguing questions about them. 

\section{Comments on the Spectral-Energy-Distribution (SED) of blazars}

Fig.\ 1 displays several examples of the `double-humped' broad-band SED, which is characteristic of blazars. 
The hump at lower frequencies, which extends at least up to the near-infrared, is uniquely identified as 
synchrotron radiation from a relativistic jet. The physical mechanism underlying the high-frequency hump is still 
debated. A popular scenario has been inverse-Compton upscattering by the jet's relativistic charged particles, 
of photons originating outside the jet (e.g., Dermer et al.\ 1992; Sikora, Begelman \& Rees 1994; Tavecchio et al.\ 
2000; Celotti et al.\ 2001; Harris \& Krawczynski 2002; Sikora et al.\ 2009 and references therein); see, however, 
Sect.\ 4. Blazar SEDs are classified in terms of the peak frequency, $\nu_{peak}$, of the `synchrotron hump'
(Abdo et al.\ 2010; also, Padovani \& Giommi 1995; Nieppola et al.\ 2006):
(i) Low-Synchrotron-Peaked (LSP): if $\nu_{peak} < 10^{14}$ Hz, (ii) Intermediate-Synchrotron-Peaked (ISP): 
$10^{14} < \nu_{peak} < 10^{15}$ Hz, and (iii) High-Synchrotron-Peaked (HSP): $\nu_{peak} > 10^{15}$ Hz.\par

\begin{figure}[h]
\centering
\includegraphics[height= 8 cm, width=12cm]{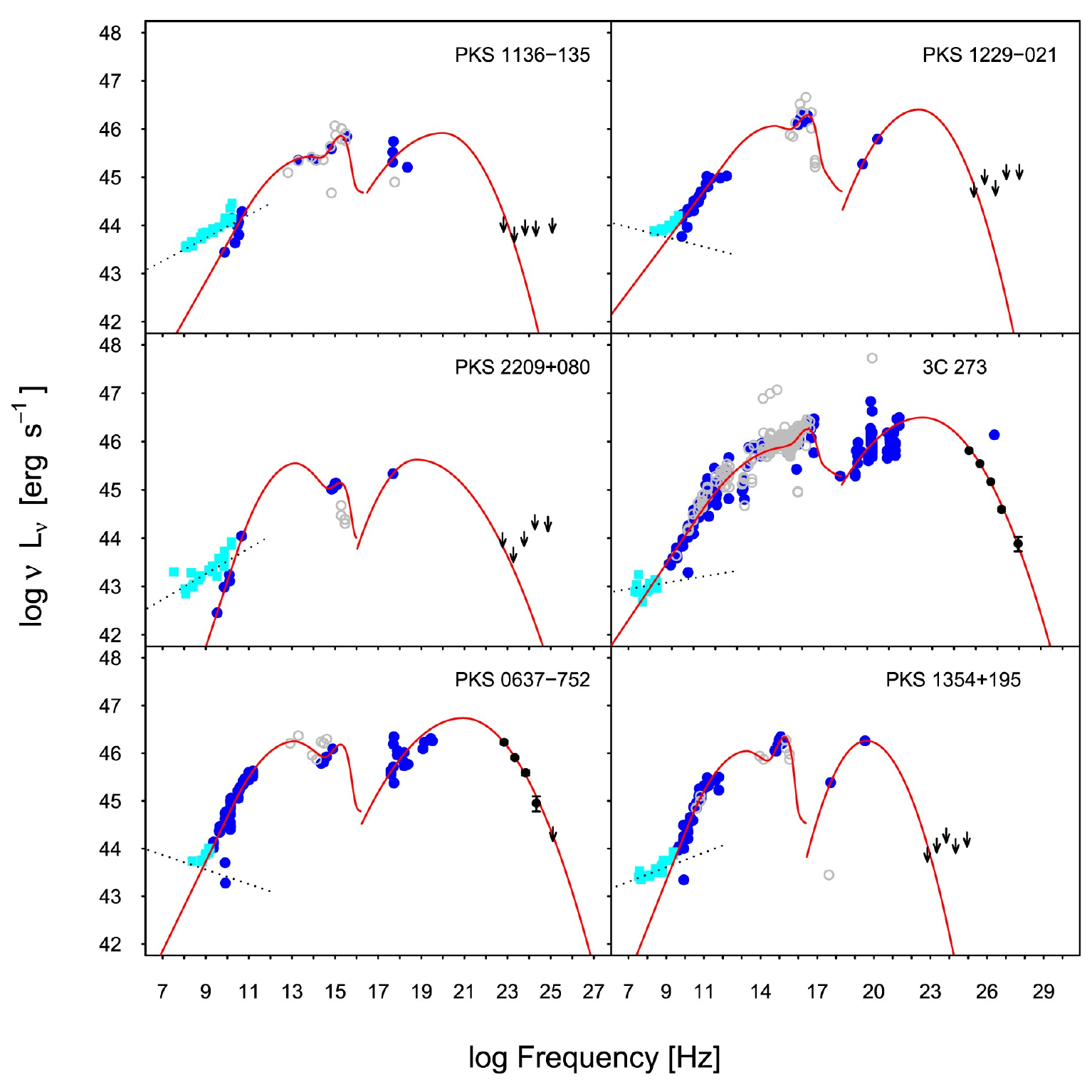}
\caption{\footnotesize{Broad-band SEDs of 6 blazars, together with their parametric models fits shown as
solid red lines (details in Breiding et al.\ 2017); the radio lobe spectrum is plotted as cyan squares, 
$Fermi$/LAT data in black, the rest of the data with blue circles and the data filtered out from the fitting 
are shown as grey circles. The non-detections at high frequencies are discussed in Sect.\ 4.2.
From Fig.\ 2 in Breiding et al.\ (2017), \copyright AAS. Reproduced by permission.} \label{fig_1}}
\end{figure}

Among blazars, SEDs of FSRQs nearly always fall within the LSP category, whereas all 3 SED types are populated
by BL Lacs (e.g., Padovani et al.\ 2017). 
Some key aspects of flux variability of AGN, in general, have been summarized recently by Smith et al.\ (2018), 
in the following terms:\par
(i) UV light tends to be more variable than the optical; (ii) variability anti-correlates with optical luminosity; 
(iii) AGN having central black-holes of highest mass $(M_{BH})$ tend to produce lower (or, slower) variability; 
(iv) the relationship of variability with the Eddington ratio is much less certain; (v) the `break time-scale' of 
the `Power Spectral Density' function correlates positively with M$_{BH}$.

\section{Do FSRQs and BL Lacs exhibit contrasting variability trends?}

An interesting clarification has come from the finding that FSRQs and BL Lacs display similar levels of 
optical variability if only radio-selected samples (i.e., the LSP type) are considered (Hovatta et al.\ 2014).
Another debated issue concerns the correlated changes between optical flux and colour. It has been claimed
that whereas for BL Lacs, colour becoming bluer with increasing brightness is more common,
the opposite happens in the case of FSRQs (e.g., Rani et al.\ 2010; Osterman Meyer et al.\ 2009; Gu et al.\ 2006, 
and references therein). This has fuelled the speculation that the jets in the two blazar species differ in some 
fundamental way. Some clarity to this conundrum has emerged from the study by Ikejiri et al.\ (2011) of 42 blazars
($m_{R} < $ 16-mag, $z$ = 0.02 - 1.75), which they monitored simultaneously in the Optical (V) and near-IR (J,K) 
bands on day/week time scales. The sample consists of 13 FSRQs, 8 LSP BLs, 9 ISP BLs and 12 HSP BLs. Their key 
conclusions are:\par

(i) Out of the 32 blazars having more than 10 flux measurements, 23 showed a clear `bluer-when-brighter' trend, at 
least in the bright state. No clear counter-example was found. \par
(ii) Some FSRQs did show 'redder-when-brighter' trend, only to revert to the `bluer-when-brighter' trend when in a 
bright state. \par
(iii) So, the `redder-when-brighter' trend displayed by some FSRQs in faint state is probably because the bluer 
thermal emission from accretion disk becomes a significant contributor to optical flux. 
{\it Thus, the colour behaviour of jet emission is indistinguishable between FSRQs and BL Lacs}.\par
Further, prominent short-term optical flares were often found to exhibit spectral hysteresis: the emission in the 
rising phase being bluer than in the decaying phase around the flare maxima. This disfavours the notion that a 
mere change in the jet's Doppler factor, e.g., due to changing jet angle, could account for the `bluer-when-brighter' 
trend (given that the Doppler boost raises both flux and frequency in tandem).\par

From these extensive data, Ikijeri et al.\ (2011) have inferred 3 components in blazars' optical emission:\par
(a) Short-term (nonthermal) flares (on days/week time scale) which exhibit spectral hysteresis.\par
(b) A significant long-term component of constant, $\it redder$ colour, varying on month/year time scale. This 
underlying component is presumably also of synchrotron origin, since thermal emission from AGN should be much bluer 
than that inferred for the long-term variable component.\par
(c) In the SEDs of some FSRQs, a blue thermal component (accretion disk) is also evident.\par
Another two interesting findings in their study are:
(i) Blazars with the synchrotron hump having a higher $\nu_{peak}$ exhibit less variability in flux, colour, and 
polarisation (see, also, Heidt \& Wagner 1998). 
(ii) In contrast to the clear correlation between optical flux and colour, the correlation between optical flux 
    and polarisation is weak.
	
\section{Is the X-ray emission from blazar jets indeed inverse-Compton?}  

Historically, a popular explanation for the X/$\gamma$-ray emission from kiloparsec-scale relativistic jets 
of powerful blazars has been inverse-Compton scattering of the CMB photons by the jet's relativistic electrons 
(IC/CMB model; Tavecchio et al.\ 2000; Celotti et al.\ 2001). 
However, this seemingly attractive model now stands contested by two relatively recent observational 
results recounted in the following subsections (also, Jester et al.\ 2006; Harris \& Krawczynski 2006; 
Hardcastle et al.\ 2016).

\subsection{New insight from the {\it SPITZER} near-Infrared imaging of the kiloparsec-scale jets of the blazars 
3C 273 and PKS 1136-135} Optical flux densities of the knots in the kiloparsec-scale jets of these blazars are found 
to be distinctly in excess over the extrapolation from the {\it SPITZER} near-infrared data points but connecting 
smoothly to the X-ray data points, thus revealing a clear spectral upturn in the optical/UV region (Fig. 2). 
This leads to the
surmise that the optical emission from these extended blazar jets is made of two components: one connecting to the 
radio and a dominant one which connects to the X-ray emission (Uchiyama et al.\ 2006; 2007). This is corraborated by 
UV measurements of the 3C 273 jet with the Hubble Space Telescope (Jester et al.\ 2007).

\begin{figure}[h]
  \centering
\hspace{0.85cm}
\includegraphics[height=2cm, width=10cm]{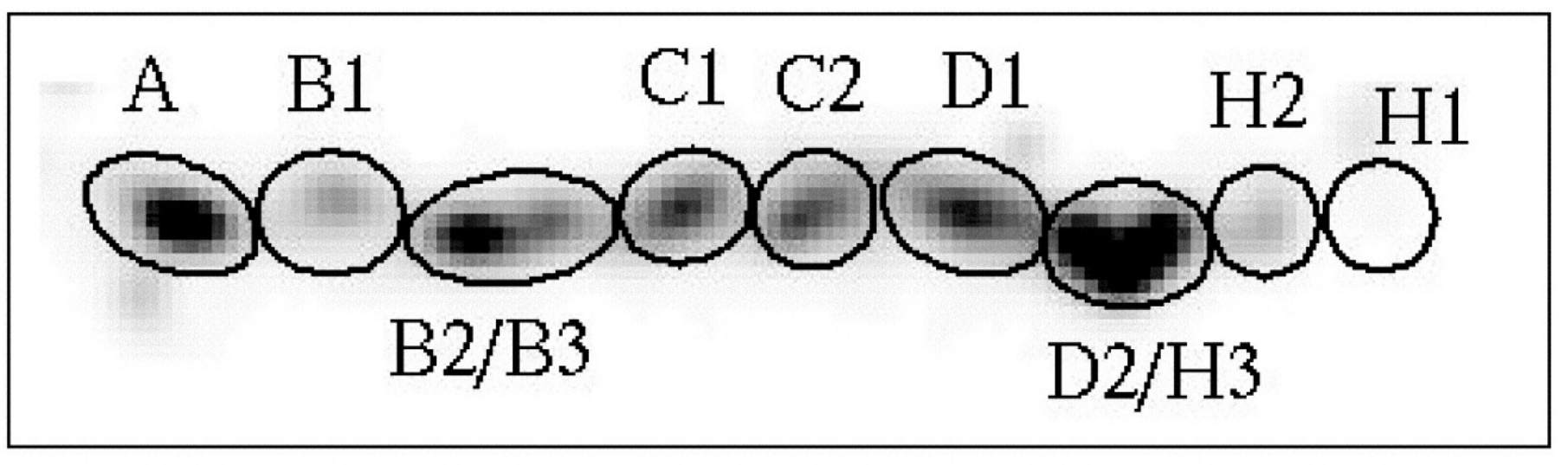}\\
\vspace{0.25cm}
\includegraphics[height=3.5cm, width=7cm]{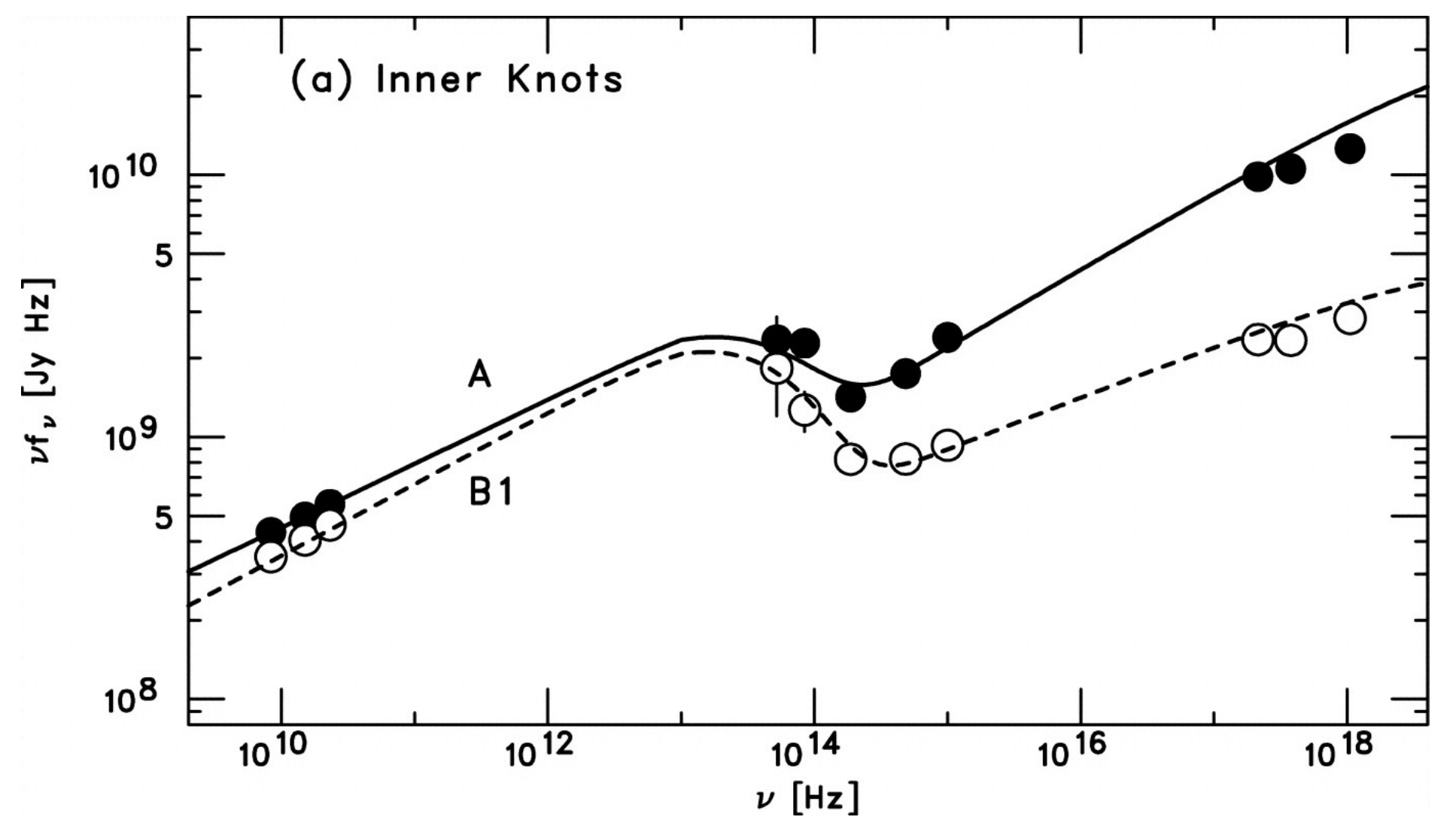}
\includegraphics[height=3.5cm, width=7cm]{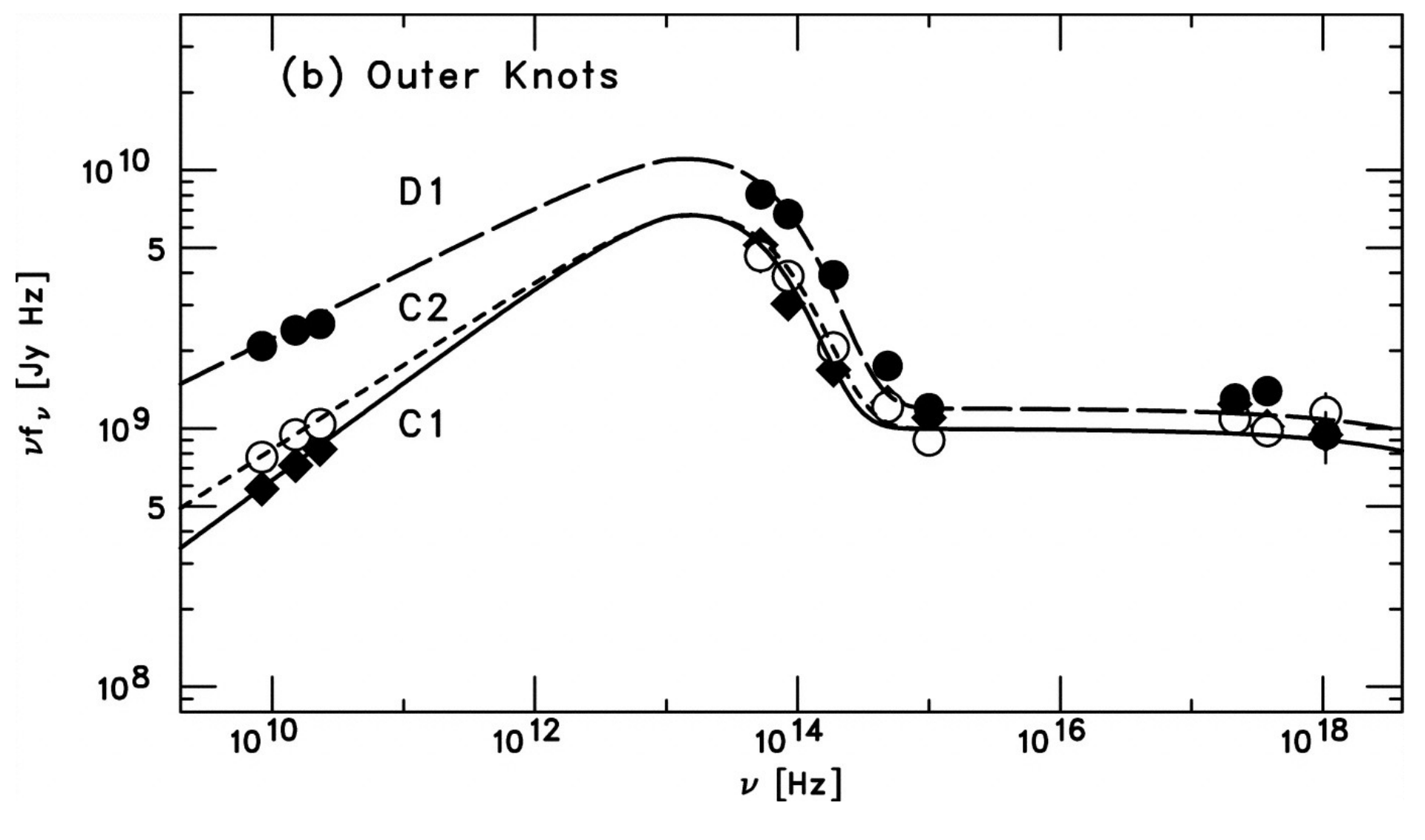}
\caption{\footnotesize{The top panel shows a {\it Hubble Space Telescope} image of the 3C273 jet, with successive knots 
marked on it. The broad-band SEDs of the knots are displayed in the lower two panels, together with the model fits.  
 From Figs.\ 4 and 5 in Uchiyama et al.\ (2006) \copyright AAS.  Reproduced with permission.}
\label{fig_1}}
\end{figure}



Quantitatively, the second spectral component is found to contribute $>$80\% of the 3C 273 jet's 
optical flux, meaning that the optical/UV extended jet largely traces its high-energy radiation 
and thus the extended jet's UV/optical emission is primarily an extension of its X-ray emission (Uchiyama 
et al.\ 2006; Jester et al.\ 2007; see, also Atoyan \& Dermer 2004; Harris et al.\ 2004; Hardcastle  2006). 
Using the Hubble Space Telescope, Cara et al.\ (2013) have measured $>$ 30\% polarization for the rising 
UV component of the extended jet of PKS 1136-135 (which is clearly linked to the high energy hump of its SED). 
This strongly suggests that the synchrotron mechanism is the dominant contributor to the jet's emission even in 
the UV/X-ray regime. 

The emerging scenario then is that the optical radiation from the extended blazar jets mainly comprises of two 
synchrotron components, one linked to the radio emission and the other to the X-ray emission. The latter 
synchrotron component underscores the need for a {\it second population} of relativistic charged particles. 
One possibility is turbulent acceleration in the shear boundary layers of the relativistic jet (Stawarz \& 
Ostrowski 2002; Rieger \& Mannheim 2002). As to the composition of the second population, both leptonic plasma 
(e.g., Atoyan \& Dermer 2004; Harris et al.\ 2004; Kataoka \& Stawarz 2005) and hadronic plasma (e.g., Mannheim \& 
Biermann 1989; Aharonian 2002; Honda \& Honda 2004; Petropoulou et al.\ 2017; Kusunose \& Takahara 2017) 
have been invoked. From an observational standpoint, some key issues are:
can these two optical synchrotron components be identified in the light curves of blazars and, secondly, are 
their INOV properties distinguishable.\par     

\subsection{Further evidence against the IC/CMB model for X-rays from blazar jets}

The IC/CMB model for kiloparsec-scale (leptonic) blazar jets is independently disfavoured, since it is often 
found to grossly over-predict GeV flux densities for their knots (Fig.\ 3). 
 
\begin{figure}[h]
\centering
\includegraphics[height= 9.5 cm, width=12cm]{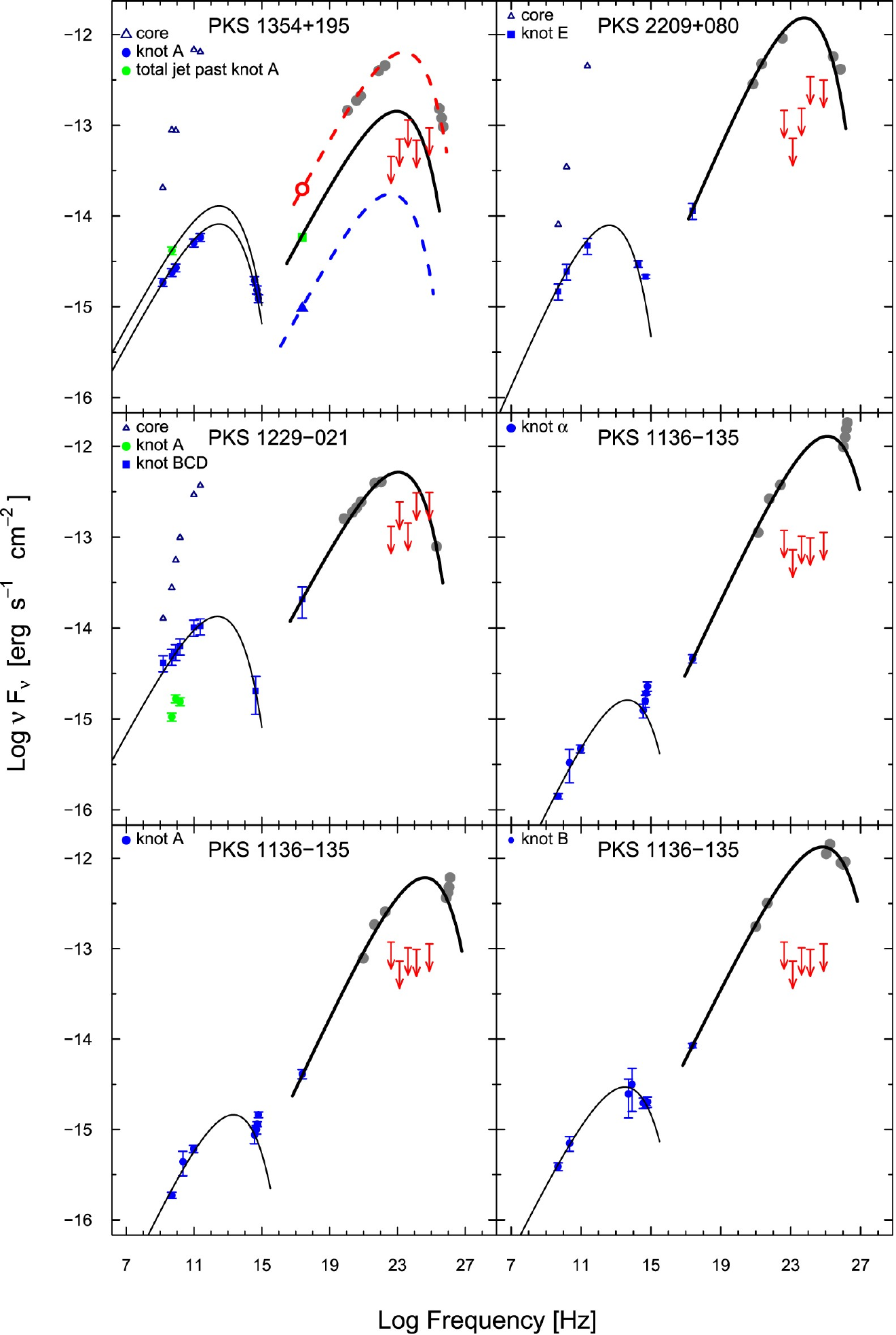}
\caption{\footnotesize{SEDs of sedgements of the kiloparsec-scale jets of four blazars. The observed data are
plotted as blue and green squares with thin solid black lines showing the synchrotron model fits to the radio-optical 
data. The curves at X-ray and $\gamma$-ray frequencies (including the grey dots) represent the IC/CMB model prediction
for the measured values at radio-optical frequencies, normalised to match the measured soft X-ray flux density (details in 
Breiding et al.\ 2017). Note the gross mismatch between the prediction and the 95\% $Fermi$/LAT upper limits shown as 
red arrows.  From Fig.\ 4 in Breiding et al.\ (2017), \copyright AAS. Reproduced by permission.}
\label{fig_1}}
\end{figure}

\section{Do optical brightness changes of AGN reflect `change of accretion state'?)}

This intriguing possibility has recently been underscored by Smith et al.\ (2018), as a straight-forward inference 
from the observed {\it bi-modality} of the optical flux distributions of a few Type I AGN (Fig. 4). Their database 
comprises of 21 light curves of Type I AGN, extracted from the {\it Kepler} archives of 4-year long optical 
monitoring data taken with a high cadence (30 min). 
Interestingly, some earlier hints of this curious behaviour were also captured in the ARIES AGN monitoring 
program (see Fig. 5 below, Stalin et al.\ 2004). The fastest known optical brightness flip has been reported for 
the TeV blazar J1555+1111, namely, a time-resolved 4\% dip, followed by return to the original level, all 
within just $\sim 1$ hour (Fig.\ 5 below, Gopal-Krishna et al.\ 2011).

\begin{figure}[h]
  \centering
\includegraphics[height=4.0cm, width=5.5cm]{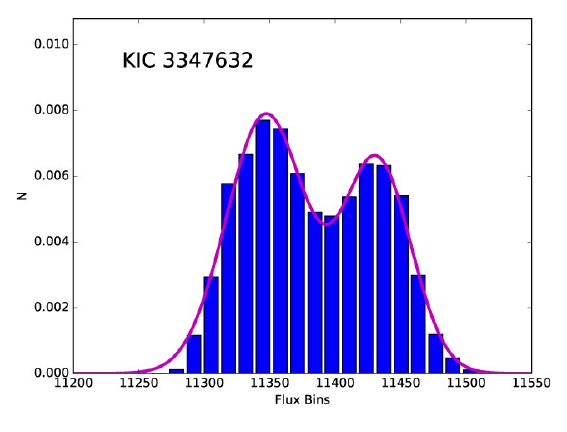}
\includegraphics[height=4.0cm, width=5.5cm]{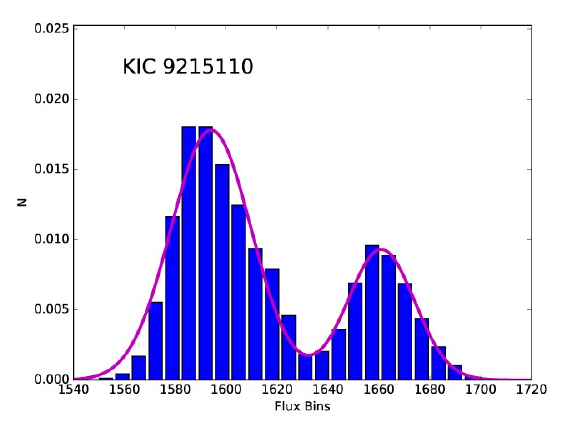}
\includegraphics[height=4.0cm, width=5.5cm]{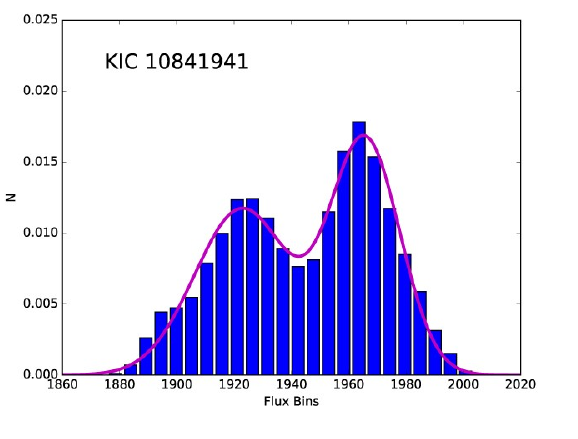}
\caption{\footnotesize{Bi-modal flux distributions of 3 Type I AGN, derived using the $Kepler$ measurements. 
 From Fig.\ 12 in Smith et al.\ (2018), \copyright AAS.  Reproduced by permission. } \label{fig_1}}
\end{figure}

\begin{figure}[h]
  \centering
\includegraphics[height=6cm, width=5.5cm]{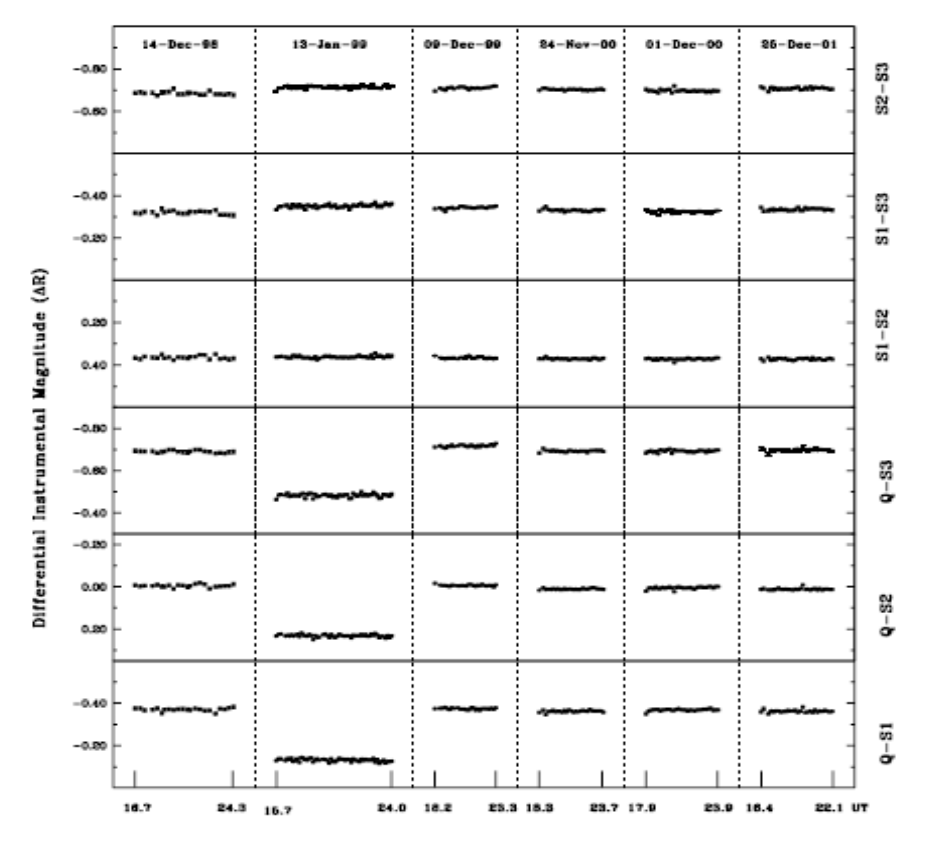}
\includegraphics[height=6.03cm, width=5.5cm]{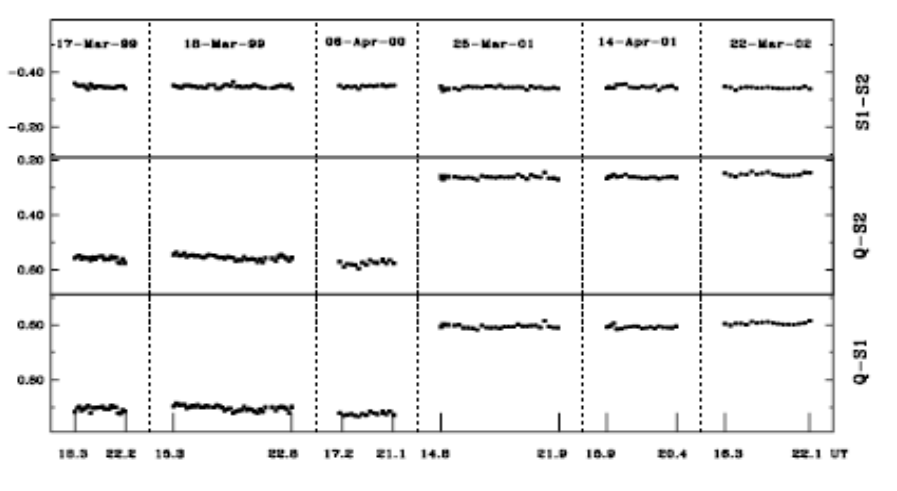}
\includegraphics[height=5.838cm, width=5.5cm]{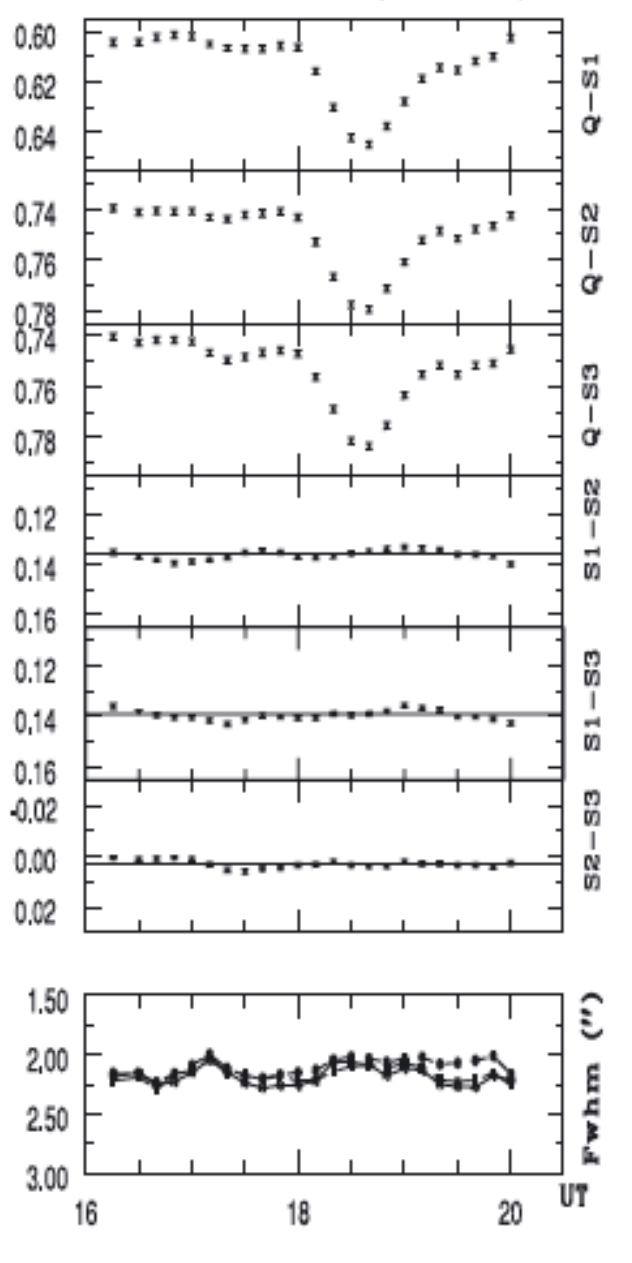}
\caption{\footnotesize{The lower 3 panels in the left figure show intranight differential light-curves (DLCs) of the 
radio-quiet QSO B0748$+$294, relative to 3 comparison stars (whose DLCs displayed in the upper 3 panels testify to 
their steadiness at all the six epochs mentioned at the top). A similar steadiness is observed for the QSO, except 
for the large ($\sim 22\%$) drop in brightness found at the second epoch. In the middle figure, the lower two panels 
present intranight DLCs for the steep-spectrum radio quasar B1103$-$006, relative to two comparison stars. The quasars 
is seen to undergo a large jump in brightness somewhere between the third and fourth epochs, while both comparison 
stars are found to remain steady through all the six epochs. All these profiles are reproduced from Figs. 5 
and 6 in Stalin et al.\ (2004), with permission from the authors. The rightmost figure displays in the top 3 panels 
the intranight DLCs of the TeV blazar J1555$+$1111, relative to 3 comparison stars. The corresponding star-star DLCs 
are shown in the lower 3 panels and these confirm the steadiness of all 3 comparison stars. The bottom panel in this 
figure displays the variation of the seeing disk through the session, which attests to steady observing conditions. 
These data are reproduced from Fig.\ 1 of Gopal-Krishna et al.\ (2011), with permission from the authors.}  \label{fig_5}}
\end{figure}

The afore-mentioned 21 {\it Kepler} light-curves of Type I AGN have also shed new light on the link between 
flux variations in the optical and X-ray bands, as highlighted by Smith et al.\ (2018):\par

(i) Unlike X-ray variability, there is no evidence for increased optical variability at higher flux.\par

(ii) Also, unlike the X-ray band, the optical flux distribution is not lognormal. This makes it unlikely
that reprocessing of X-ray flares in the AGN is a major contributor to their optical variability.
  

\section{Some highly unusual optical flaring events of AGN}

Among the afore-mentioned {\it Kepler} light curves of Type1 AGN, Smith et al.\ (2018) have noticed an unusual flare 
lasting just a few days, whose nature is unclear (hence they term it a `mysterious phenomenon', see Fig. 6 left).
They argue that the flare's duration is too short for a supernova afterglow and, furthermore, its exponential 
decline is inconsistent with the chracteristic $t^{-5/3}$  profile associated with a star's tidal disruption by 
the central supermassive black hole (e.g., Komossa 2015) . 

\begin{figure}[h]
  \centering
  \includegraphics[height=7cm, width=6cm]{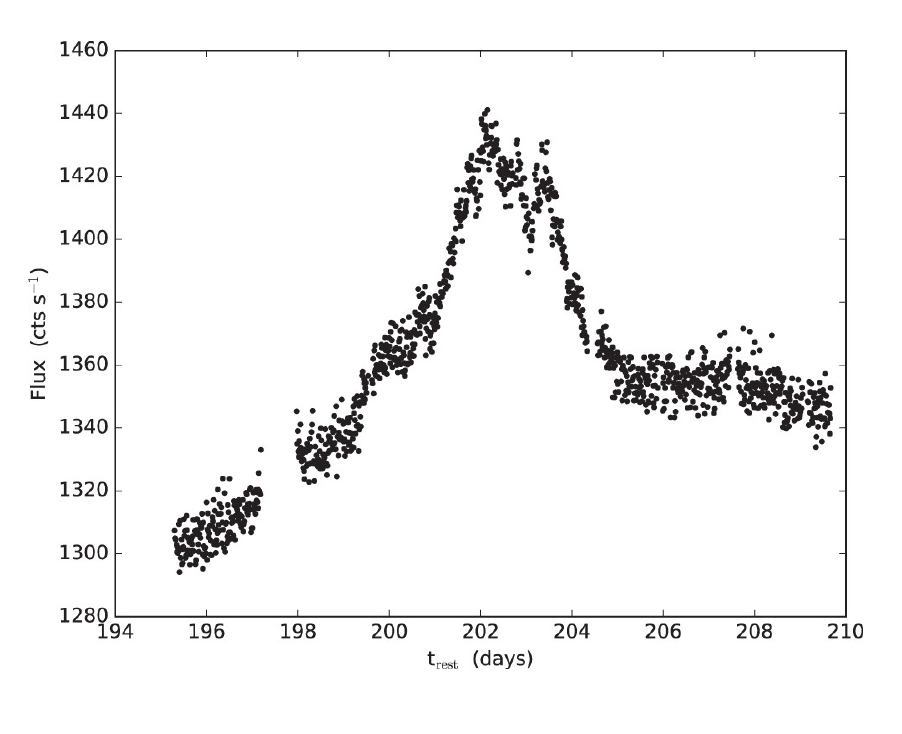}
  \hspace{1.0 cm}
  \includegraphics[height=6.858cm, width=4.0cm]{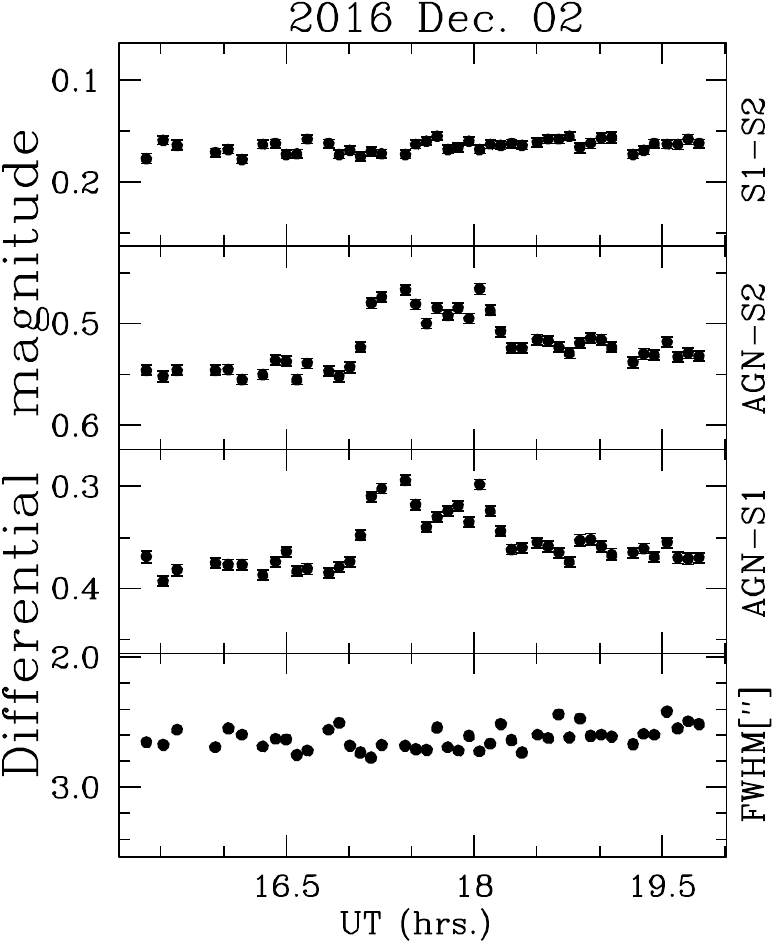}
  \hspace{.5 cm}
\includegraphics[height=7.00cm, width=4.0cm]{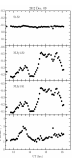}

\caption{\footnotesize{The 2-week long {\it Kepler} light-curve of the Type I AGN KIC 11606852, shown in the 
left figure is dominated by a large optical flare; from Fig.\ 22 in 
Smith et al.\ (2018), \copyright AAS. Reproduced by permission.
The other two figures present, for the $\gamma$-ray detected narrow-line Seyfert1 galaxy 1H 0323+342, 
intranight differential light curves (DLCs) drawn relative to two comparison stars (whose star-star DLCs are shown 
in the top panel of each figure). The seeing disk variation is plotted in the bottom panel (for details, see
Ojha et al.\ 2019). Reproduced with permission of the authors. The data 
in the right figure (2012-Dec-09) are from Fig.\ 10 in Paliya et al.\ (2014), \copyright AAS. Reproduced with permission.  
Note that the precursor bump seen in both light curves between 13-14 UT is an artefact of a transient deterioration of 
the seeing disk, as evident from the bottom panel.} \label{fig_6}}
\end{figure}

Here we draw attention to two more optical flares with remarkably similar appearances, albeit lasting just about an 
hour (Fig. 6). Both were recorded independently in the intranight optical minotoring of the $Fermi$/LAT 
detected narrow-line Seyfert1 galaxy (NLSy1) 1H 0323+342, 
from Hanle and Devsthal observatories in 2012 (Paliya et al.\ 2014) and 2016 (Ojha et al.\ 2019). The order-of-magnitude 
shorter durations of these two flares could be understood if NLSy1 galaxies indeed harbour much lighter black holes 
than normal AGN which are hosted by massive early-type galaxies (e.g., Komossa et al.\ 2006; Zhou et al.\ 2006). At 
least the 2012 flare is known to have occured during a high $\gamma$-ray activity state of this NLSy1 galaxy 
(Paliya et al.\ 2014). By modelling the braod-band SED using a `one-zone leptonic jet' 
model, these authors posit that over a large range of $\gamma$-ray activity level, the optical/UV 
component is dominated by thermal emission from the accretion disk, as also concluded by Zhou et al.\ (2007). 
This is not unexpected since NLSy1s galaxies are believed to be usually accreting near the Eddington rate (e.g., 
Komossa et al.\ 2006; Zhou et al.\ 2006). As pointed out by Ojha et al.\ (2019), an interesting corrollary to all 
this is that the observed rapid optical variation (which must be associated with the synchrotron jet) gets much 
diluted by the (quasi-steady) $\it thermal$ optical emission arising from the accretion disk (and the host galaxy), 
so that at most 25\% of the optical emission in the light-curve is nonthermal.
This further means that the observed 7\% increase in the {\it total} optical flux in $<$ 20 minutes (Fig.\ 6) would 
actually need a $>$ 27\% jump in the nonthermal (synchrotron) light within this short duration and that would 
correspond to a flux doubling time of just $\sim 1$ hour for the optical synchrotron emission, even under somewhat 
conservative assumptions (Ojha et al.\ 2019). A very similar conclusion has been reached by them for the optical flare 
of this NLSy1 galaxy observed on 2012-Dec-9 (Fig.\ 6). Clearly, both these outbursts exemplify extreme optical 
variability, at least an order-of-magnitude more violent than is typical for blazars (e.g., Ferrara et al.\ 2001). 
A less conservative analysis of the data could perhaps place such optical flaring events virtually at par with the 
ultra-rapid TeV flares with flux doubling time of just $\sim$ 10 minutes, as witnessed for some blazars, like PKS 
2155-304 (Aharonian et al.\ 2007) and PKS 1222+211 (Aleksic et al.\ 2011). 

\section{An intriguing polarization flare of the blazar 3C 454.3}

Rapid variation in the intensity and position angle of linearly polarized emission from blazars can unravel
crucial information on the kinematic of their relativisitic jets (e.g., Wagner \& Witzel 1995; Marscher 
et al.\ 2016). An intriguing optical polarization flare was recorded during the monitoring of the blazar 3C 454.3
on Dec., 3-12 (2009) (Fig.\ 7 below, Gupta et al.\ 2017). The reported light curves span $\sim$ 7 
months and cover radio, optical, near-infrared, X-ray and $\gamma$-ray bands. The strong flare peaked almost 
simultaneously in all the wavebands, however a radio counterpart was conspicuous by its absence. This might, in
fact, be consistent with the proposal that a different population of synchrotron emitting particles is 
responsible for the radio emission in blazars (Sect. 4.1).  

\begin{figure}[h]
 \centering
\includegraphics[height=10.0cm, width=16cm]{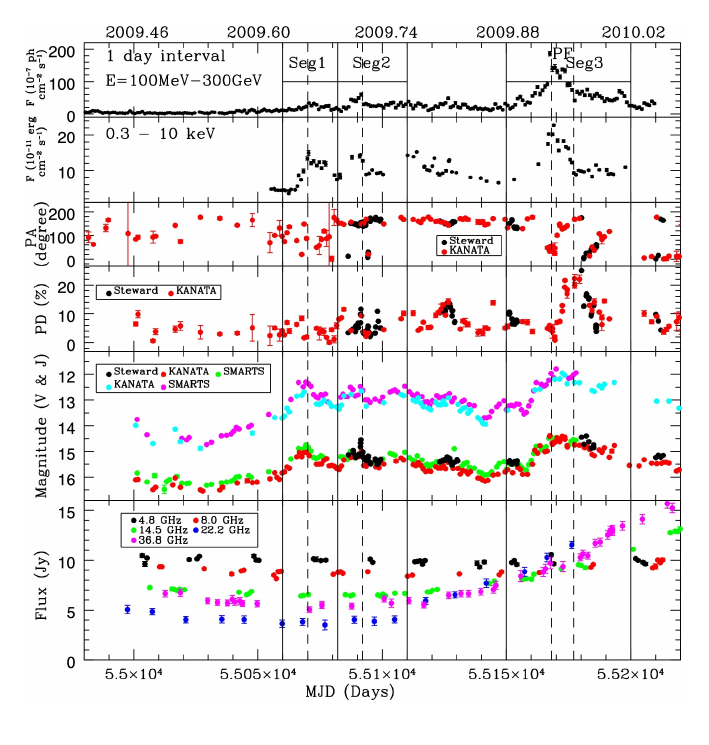}
\caption{\footnotesize{Light-curves of the blazar 3C 454.3 at radio, optical, near-infrared, X-ray and $\gamma$-ray bands, 
covering a time span of $\sim$ 7 months during 2009-10.  A large flare lasting about 10 days occured in December 2009, 
but had no radio counterpart. The associated large variation of the optical linear polarization (both in amplitude and 
position angle) are also displayed (reproduced from Fig.\ 1 of Gupta et al.\ (2017), with permission of the authors).}
\label{fig_1}}
\end{figure}

Another intriguing aspect is that the onset of the dramatic rise in the optical polarization (and its position angle) 
from  $\sim$ 3 to 20\% coincides rather precisely with the beginning of the flare's decline in the high energy bands, 
as well as in the optical/near-infrared (Fig. 7). While this striking anti-correlation might be understood in terms of 
the `swinging jet' model (Gopal-Krishna \& Wiita 1992; also, Bachev 2015), the lack of a radio counterpart points towards 
a more complex situation.

\section{New insight on `stationary' compact radio knots in blazar jets}

VLBI monitoring of blazars has revealed several examples of `stationary' radio knots whose {\it separation} from the 
core does not appear to change, the BL Lac object S5 1803+784 being a prime example of this phenomenon (Britzen et al.\ 
2010a). For another prominent BL Lac object, OJ 287, the recently collated exceptionally dense VLBI monitoring data are 
particularly revealing. 

\begin{figure}[h]
  \hspace{1.5 cm}
  \includegraphics[height=5.0cm, width=7.5cm]{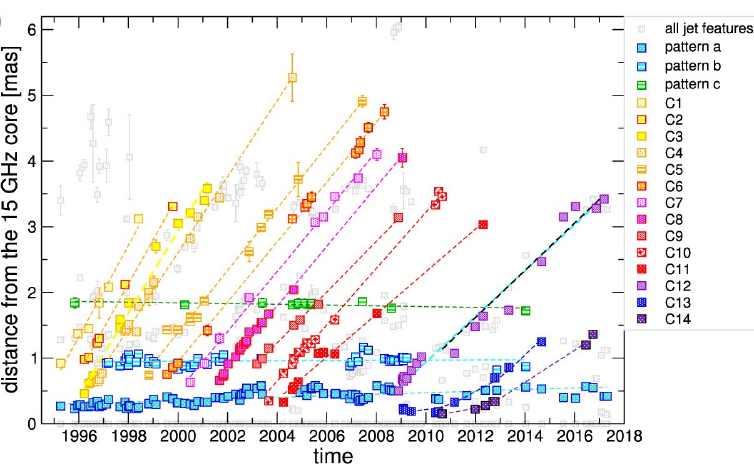}
  \hspace{0.5 cm}
\includegraphics[height=5.0cm, width=6.5cm]{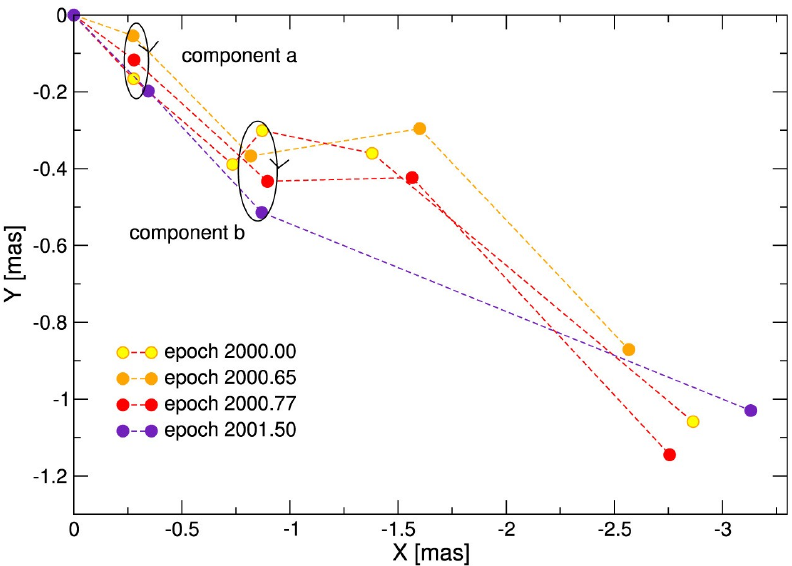}

\caption{\footnotesize{The left figure is a plot of distance from the 15 GHz core as a function of time, for the 17 VLBI 
radio knots identified in the nuclear jet of the blazar OJ 287. Three of these knots are seen to maintain constant distance
from the core and hence they could be (naively) termed `stationary' knots (as they do not participate in the global outward 
motion). For the innermost two of the 3 knots, the right figure shows the trajectory in the 2-dimensional sky plane, 
measured at 4 epochs between 2000.0 and 2001.5. Both these radio knots appear to trace a loop on a year-like time scale 
(reproduced from Figs.\ 4a and 5a in Britzen et al.\ (2018), with permission from the authors).}  
\label{fig_1}}
\end{figure}

Analysis of 120 VLBI observations of OJ 287 at 15 GHz, over the period 1995-2017, has enabled tracking the proper motions 
of its 17 radio knots, out of which 3 are found to be `stationary' (i.e., maintaining a constant separation from the 
core), while the remaining 14 knots exhibit steady outward motions (Fig. 8). Tracing of the motion in 
the 2-dimensional space, however, reveals that {\it even the `stationary' knots move}. Modelling of their motions
(Britzen et al.\ 2018) shows that they may well be tracing the nuclear relativistic jet's rotation/nutation on time scale 
of $\sim$ 1 year. Several examples of `stationary' VLBI knots actually performing apparent transverse motion have been 
documented in the literature (e.g., Britzen et al.\ 2010a,b; Kun et al.\ 2018).



\section{Conclusions}
For blazar type AGNs featuring prominent, relativistically boosted nonthermal jets, we have highlighted 
a number of observational results drawn from the relatively recent literature. These have unveiled striking, 
often curious, aspects of their emission processes and temporal variability across the observable 
electromagnetic spectrum. The selection of these results by us is by no means exhaustive, or unbiased. It is only 
meant to turn a spotlight on a cross-section of observations (some of which have received scant attention) that 
have either helped in clarifying certain long standing issues about the functioning of AGN, or have posed some 
new intriguing questions whose sustained follow-up could make an important contribution to the consolidation of 
theoretical models of the AGN phenomenon. The 3-metre class optical telescopes stationed at Devsthal can play a 
very useful role in this quest.

\section{Acknowledgements}

We thank Mr. V. Ojha for help with the literature search.

\footnotesize
\beginrefer







\refer Abdo A. A., Ackermann M., Ajello M. et al. 2010, ApJ, 716, 30

\refer Aharonian F. A. 2002, MNRAS, 332, 215

\refer Aharonian F., Akhperjanian A. G., Bazer-Bachi A. R. et al. 2007, ApJL, 664, L71

\refer Aharonian F. A., Barkov M. V., Khangulyan D. 2017, ApJ, 841, 61

\refer Aleksic J., Ansoldi S., Antonelli L. A. et al. 2014, Sci, 346, 1080

\refer Aleksic J., Antonelli L. A., Antoranz P. et al. 2011, ApJ, 730, L8

\refer Antonucci R. 2012, A\&AT, 27, 557

\refer Bachev R. 2015, MNRAS, 451, L21

\refer Begelman M. C., Blandford R. D.,  Rees M. J. 1984, RvMP, 56, 255

\refer Breiding P., Eileen T. M., Georganopoulos M. et al.\ 2017, ``Fermi Non-detections of Four X-ray Jet Sources and Implications for the IC/CMB Mechanism'', ApJ, 849, 95; doi: 10.3847/1538-4357/aa907a

\refer Britzen S.; Kudryavtseva N. A.; Witzel A. et al. 2010a, A\&A, 511A, 57

\refer Britzen S., Witzel A., Gong B. P. et al. 2010b, A\&A, 515A, 105

\refer Cara M., Perlman E. S., Uchiyama Y. et al. 2013, ApJ, 773, 186

\refer Celotti A., Ghisellini G., Chiaberge M. 2001, MNRAS, 321, L1

\refer Dermer C. D., Schlickeiser R., Mastichiadis A. 1992, A\&A, 256, L27

\refer Ferrara E. C., Miller H. R., McFarland J. P. et al. 2001, ASPC, 224, 319

\refer Georganopoulos M., Perlman E. S., Kazanas D. 2006, ApJL, 653, L5 

\refer Gopal-Krishna, Stalin C. S., Sagar R. et al. 2003, ApJL, 586, L25

\refer Gopal-Krishna, Wiita P. J., 1992, A\&A, 259, 109

\refer Gopal-Krishna, Wiita, P. J. 2018, BSRSL, 87, 281

\refer Gu M. F., Lee C.-U., Pak S. et al. 2006, A\&A, 450, 39

\refer Gupta A. C., Mangalam A., Wiita P. J. et al. 2017, MNRAS, 472, 788

\refer Hagen-Thorn V. I., Larionov V. M., Jorstad S. G. et al. 2008, ApJ, 672, 40 

\refer Hardcastle M. J. 2006, MNRAS, 366, 1465

\refer Hardcastle M. J., Lenc E., Birkinshaw M. et al. 2016, MNRAS, 455, 3526

\refer Harris D. E., Krawczynski H. 2002, ApJ, 565, 244

\refer Harris D. E., Krawczynski H. 2006, ARA\&A, 44, 463

\refer Harris D. E., A. E. Mossman, Walker R. C. 2004, ApJ, 615, 161

\refer Heidt J., Wagner S. J. 1998, A\&A, 329, 853

\refer Honda Y. S., Honda, M. 2004, ApJL, 613, L25 

\refer Hovatta T., Pavlidou V., King O. G. et al. 2014, MNRAS, 439, 690

\refer Ikejiri Y., Uemura M., Sasada M. et al. 2011, PASJ, 63, 639

\refer Jester S., Harris D. E., Marshall H. M. et al. 2006, 648, 900  

\refer Jester S., Meisenheimer K., Martel A. et al. 2007, MNRAS, 380, 828

\refer Kataoka J., Stawarz L. 2005, ApJ, 622, 797

\refer Komossa S. 2015, JHEAp, 7, 148

\refer Komossa S., Voges W., Xu D. 2006, AJ, 132, 531


\refer Kun E., Karouzos M., Gabanyi K. E. 2018, MNRAS, 478, 359

\refer Kusunose M., Takahara F. 2017, ApJ, 835, 20

\refer Mannheim K., Biermann P. L. 1989, A\&A, 221, 211S

\refer Narayan R., Piran T. 2012, MNRAS, 420, 604

\refer Nieppola E., Tornikoski M., Valtaoja E. 2006, A\&A, 445, 441

\refer Ojha V., Gopal-Krishna, Chand H. 2019, MNRAS, 483, 3036	

\refer Osterman Meyer A., Miller H. R., Marshall K. 2009, AJ, 138, 1902

\refer Padovani P., Alexander D. M., Assef R. J. et al. 2017, A\&ARv, 25, 2

\refer Padovani P., Giommi P. 1995, ApJ, 444, 567

\refer Paliya V. S., Sahayanathan S., Parker M. L. et al. 2014, ``The Peculiar Radio-loud Narrow Line Seyfert 1 Galaxy 1H 0323+342'', ApJ, 789, 143; doi: 10.1088/0004-637X/789/2/143 

\refer Petropoulou M., Nalewajko K., Hayashida M. et al. 2017, MNRAS, 467, L16 

\refer Rani B., Gupta A.~C., Strigachev A. et al. 2010, MNRAS, 404, 1992

\refer Rieger F. M., Mannheim K. 2002, A\&A, 396, 833

\refer Sikora M., Begelman M.~C., Rees M.~J. 1994, ApJ, 421, 153

\refer Sikora M., Stawarz L., Moderski R. et al. 2009, ApJ, 704, 38

\refer Smith K. L., Mushotzky R. F., Boyd, P. T. et al. 2018, ``The Kepler Light Curves of AGN: A Detailed Analysis'', ApJ, 857, 141; doi: 10.3847/1538-4357/aab88d

\refer Tavecchio F., Maraschi L., Sambruna R. M. et al. 2000, ApJL, 544, L23

\refer Uchiyama Y., Urry C.~M., Cheung C.~C. et al.\ 2006, "Shedding New Light on the 3C 273 Jet with the Spitzer Space Telescope'', ApJ, 648, 910-921; doi: 10.1086/505964

\refer Uchiyama Y., Urry C.~M., Coppi P. et al. 2007, ApJ, 661, 719

\refer Urry C. M., Padovani P. 1995, PASP, 107, 803

\refer Zhou H., Wang T., Yuan W. et al. 2006, ApJS, 166, 128

\refer Zhou H., Wang T., Yuan W. et al. 2007, ApJL, 658, L13

\endrefer           

\end{document}